%% file: lns_ext.tex
\documentclass[copyright,creativecommons]{eptcs}
\providecommand{\event}{ICE 2022} % Name of the event you are submitting to

\usepackage{breakurl}             % Not needed if you use pdflatex only.
\usepackage{underscore}           % Only needed if you use pdflatex.

\usepackage{booktabs}   %% For formal tables:
\usepackage{stmaryrd}

\usepackage{amssymb,amsmath,amsthm}
\usepackage{color}
\newtheorem{definition}{Definition}\usepackage{color}
\usepackage{color}

\usepackage{listings}
\usepackage{mathpartir}
\usepackage{pifont}

\usepackage{tcolorbox}
\usepackage{tabularx}
\usepackage{xcolor}
\usepackage{wrapfig}
\usepackage{bm}

\usepackage{float}
\usepackage[sort]{cite} 

\usepackage{centernot}

\input{definitions.tex}

\usepackage{semantic}

\title{Lang-n-Send Extended: Sending Regular Expressions to Monitors}
\author{Matteo Cimini
\institute{University of Massachusetts Lowell \\ Lowell, MA, USA}
\email{matteo\_cimini@uml.edu}
}
\def\titlerunning{Lang-n-Send Extended: Sending Regular Expressions to Monitors}
\def\authorrunning{M. Cimini}
\begin{document}
\maketitle

\begin{abstract}
In prior work, Cimini has presented $\textsc{Lang-n-Send}$, a $\pi$-calculus with language definitions. 

In this paper, we present an extension of this calculus called $\textsc{Lang-n-Send}^{\texttt{+m}}$. 
First, we revise $\textsc{Lang-n-Send}$ to work with transition system specifications rather than its language specifications. 
This revision allows the use of negative premises in deduction rules.   
Next, we extend $\textsc{Lang-n-Send}$ with monitors and with the ability of sending and receiving regular expressions, which then can be used in the context of larger regular expressions to monitor the execution of programs. 

We present a reduction semantics for $\textsc{Lang-n-Send}^{\texttt{+m}}$, 
and we offer examples that demonstrate the scenarios that our calculus captures. 
\end{abstract}

\section{Introduction}\label{introduction}

As the field of software language engineering advances \cite{LangWorkbenches}, 
it is increasingly easier for programmers to quickly define and deploy their own programming languages. 
Cimini has presented in \cite{lns} a $\pi$-calculus called $\lnsOLD$ that accommodates \quoting{language-oriented} concurrent scenarios. 
Processes of $\lnsOLD$ can define languages with a syntax for structural operational semantics (SOS), and use these languages to execute programs. 
Processes can also send and receive fragments of operational semantics through channels. 
An example of the type of scenarios that $\lnsOLD$ enables is the following, which is a simplified version of the first example in \cite{lns}.  
%To provide our readers with an intuition as to what type of scenarios $\lnsOLD$ enables, let us consider a simplified version of the first example in \cite{lns}. % following. 
A server stores a library of operational semantics definitions. 
This server can send these fragments of semantics through a channel upon requests. 
At a certain point, a client defines a program and a language with which it intends to execute the program. 
The client notices that the program that it is about to execute is safety-critical. 
Therefore, the client would like to execute the program in the context of the disrupt operator \cite{Baeten00modetransfer}. 
The client, then, requests the fragment of operational semantics that defines the disrupt operator from the server, and receives it through a channel. 
The client augments its language with the disrupt operator, and executes the program using this new language. 
In particular, it uses the disrupt operator to also specify the error code to be executed in case of a disruption. 

The crux of $\lnsOLD$ consists of two operators: 
Program executions $\execLNN{\mathcal{L}}{x}{}{\textit{program}}{\textit{trace}}$ and $\isInTraceOpOLD$, used as follows: 

\[
\execLNN{\mathcal{L}}{x}{}{\textit{program}}{\textit{trace}} \parSym x(\textit{trace}).\isInTraceOLD{a}{\textit{trace}}{P}{Q}
\]

where $\mathcal{L}$ is a language definition and $x$ is a channel. $\execLNN{\mathcal{L}}{x}{}{\textit{program}}{\textit{trace}}$ uses the operational semantics of $\mathcal{L}$ to prove a transition from \textit{program}. 
The evaluation of \textit{program} proceeds this way one step at a time. 
Each transition is labelled, and $\lnsOLD$ accumulates the execution trace in $\textit{trace}$. 
When \textit{program} terminates, the final trace is sent over the channel $x$. 
The process on the right of the parallel operator receives the trace, and analyzes it with $\isInTraceOLD{a}{\textit{trace}}{P}{Q}$. 
This process checks that the label $a$ is one of the labels in \textit{trace}, and continues as $P$ in such a case. Otherwise, it continues as $Q$. 

This paper addresses two limitations of $\lnsOLD$.

\paragraph{First Addition: from Higher-order Logic Programs to Transition System Specifications}

\cite{lns} provides a syntax for language definitions $\mathcal{L}$. 
This syntax has been specifically devised to represent operational semantics.
The semantics of $\mathcal{L}$ is based on higher-order logic programming as realized with hereditary Harrop formulae \cite{Miller:2012lp}: 
$\mathcal{L}$ is compiled into a higher-order logic program $\mathcal{P}$, 
and $\lnsOLD$ computes the steps of \textit{program} using the provabilty relation $\models$ of higher-order logic programming \cite{Miller:2012lp}, i.e., $\mathcal{P} \models (\step \app \textit{label} \app \textit{program} \app \textit{program}')$, for some $\textit{program}'$ and $\textit{label}$. 

However, higher-order logic programs of \cite{Miller:2012lp}, and therefore $\models$, do not contemplate the use of negation. %, which is generally unsound in $\lambda$-prolog. 
This prevents $\lnsOLD$ from defining languages with operators that use negative premises. 

In this paper, we revise $\lnsOLD$ to adopt transition system specifications (TSSs) \cite{Bol1996}, a well-known and widely used formalism for SOS specifications. 
Our motivation for adopting TSSs is that they include negative premises with a well-established semantics \cite{Groote1992}. 
This is a consequential addition: It has been shown that negative premises are actually necessary to express some operators such as the priority operator \cite{aceto2008}. 
That is, SOS with negative premises is strictly more expressive than SOS without them. 
Other examples of operators that use negative premises include timed operators \cite{tpa}, and some formulations of sequential composition, to name a few. 

Our salient challenge here consists in incorporating TSSs into $\lnsOLD$. 
Fortunately, the design of $\lnsOLD$ is rather {modular} insofar this aspect is concerned, and we simply can use the definitions of TSSs from the literature \cite{Groote1992,Bol1996,vanGlabbeek2004} in lieu of the syntax for language definitions $\mathcal{L}$.

\paragraph{Second Addition: Online Monitors and Communication of Regular Expressions} 
Monitoring is a runtime verification technique that is based on executing a program and observing its behavior. 
Its goal is to establish whether such execution satisfies or violates a correctness property 
(see \cite{monitorSurvey} for a survey on the subject).
%An excellent survey on monitoring is \cite{monitorSurvey}. 
There are two types of monitoring: \emph{offline} monitoring and \emph{online} monitoring. 
Offline monitoring executes the program and records its execution trace. 
The trace is then analyzed after the execution terminates. 
Conversely, online monitoring performs its analysis alongside the execution of the program.  
That is, an online monitor acts after each step of the execution, and analyzes the trace that has been generated up to that point. 

$\lnsOLD$ includes offline monitoring with $\isInTraceOpOLD$. 
This operation only checks whether an action has occurred. 
However, this is an inexpressive form of trace analysis. 
For example, suppose that we were to check whether a process performs valid actions on a file. 
We would need to see whether read/write operations are performed \emph{after} opening the file. 
We would also need to check whether the file is being closed at the end. 
Simply checking whether an action has occurred with $\isInTraceOpOLD$ is not enough, as we need to express more refined temporal properties. 

Clearly, $\isInTraceOpOLD$ is insufficient for most scenarios. 
Therefore, this paper extends $\lnsOLD$ with a more powerful way of analyzing traces. 
Specifically, we augment $\lnsOLD$ with regular expressions, and we add the ability of checking whether a trace satisfies or violates a regular expression. 
Our salient challenge here consists in equipping $\lnsOLD$ with appropriate linguistic features for the monitoring of regular expressions. 
We do that as follows.

$\lnsOLD$ does \emph{not} feature online monitoring. We then model it with an extended form for program executions: 
\[
 \execLNNMonitor{\TSS}{x}{}{\textit{program}}{\textit{trace}}{}{}{m_1 ~ m_2 ~ \cdots ~ m_n}
\]

where $\TSS$ is a TSS and $m_1$, $m_2$, $\ldots$, and $m_n$ are online monitors. Each $m_i$ carries the regular expression to be checked during the execution of \textit{program}, as well as a process to be executed in case such regular expression is violated.

We also have extended the offline monitoring capabilities of $\lnsOLD$ by replacing $\isInTraceOpOLD$ with the following operation: 
\[\isInTrace{\textit{trace}}{\textit{regexp}}{P}{Q}\]

This operation checks that the regular expression \textit{regexp} validates \textit{trace}. If the check is successful then we continue with $P$, otherwise we continue with $Q$. 
(Our $\isInTraceOp$ operator is not limited to checking acceptance for regular expressions, though we postpone this discussion to Section \ref{lns}.) 
%For reasons that we explain later in the paper, $\isInTraceOp$ is not limited to check acceptance for regular expressions. We will discuss this aspect in Section \ref{lns}.  

$\lnsOLD$ is tailored to express dynamic scenarios where language fragments are sent and received, 
and where processes are instructed to execute programs received from other processes. 
In this dynamic context, it is natural to also receive, from external processes, the regular expressions to monitor. 
We have therefore extended $\lnsOLD$ with the ability of sending and receiving regular expressions through channels. 

We call this new calculus $\lns$ (as in \quoting{plus monitoring}). 
Notice that, in this paper, our goal is to address the operations described above with suitable linguistic features. We argue in Section \ref{discussion} that a calculus that is smaller than the one that we show, and that can encode our operations, may exist, but we leave exploring that research venue as future work.

\paragraph{Contributions}

We present a reduction semantics for $\lns$ in Section \ref{operationalSemantics}. 
To demonstrate the type of scenarios that $\lns$ captures, we provide the following examples in Section \ref{examples}:

\begin{itemize}
\item \textbf{Negative premise (Example 1).} A server receives the semantics of the parallel operator from another process, which decides whether parallel processes are allowed to spend idle time or whether they must run with maximal progress. 
\item \textbf{Offline monitoring (Example 2).} A server receives programs from clients, executes them to the end, and checks that the programs have used files correctly (open before read/write operations, and close at the end). The server does so by checking that the final trace is accepted by an appropriate regular expression. 
\item \textbf{Online monitoring and sending/receiving of regular expressions (Example 3).} This example refines Example 2. 
Programs can perform a privileged action on files as long as they respect a correct sequence of actions. 
This sequence of actions changes every day, and is provided by an external process. 
The server receives this sequence as a regular expression, and uses it to install an online monitor for the execution of programs. 
\end{itemize}

We believe that $\lns$ provides a suitable formalism to express these and similar scenarios. % a first step to a firm foundation for this type of programming. 
The paper is organized as follows. Section \ref{syntaxLanguage} provides the definition of TSSs from the literature. % transition system specifications
Section \ref{lns} presents the syntax of $\lns$. 
Section \ref{operationalSemantics} presents a reduction semantics for $\lns$.  
Section \ref{examples} demonstrates our calculus with the examples described above. 
Section \ref{discussion} offers a discussion of selected aspects such as deadlocks, implementation ideas, and ideas for simplifying our calculus. 
Section \ref{relatedWork} discusses related work, and 
Section \ref{conclusion} concludes the paper.

\section{Preliminaries: Transition System Specifications}\label{syntaxLanguage}

We recall the definitions for transition system specifications from \cite{Groote1992,Bol1996}. 

\begin{definition}[Signatures and Terms]\label{signatures}
A \emph{signature} $\Sigma$ is a pair $(F, \textit{ar})$ where $F$ is a set of function symbols, and the function $\textit{ar} : F \to \mathbb{N}$ determines the arity of the functions in $F$.  
Given a signature $\Sigma= (F, \textit{ar})$, $T(\Sigma)$ is the \emph{set of terms of the signature} $\Sigma$, and is defined as the minimal set satisfying the following: \\
(We use the symbol $t$ for terms). 

\begin{itemize}
\item $\varSet \subseteq T(\Sigma)$, where $\varSet$ is a set of variables, 
\item if $t_1, \ldots,t_n \in T(\Sigma)$, $f\in F$, and $\textit{ar}(f) = n$ then $f(t_1, \ldots,t_n) \in T(\Sigma)$. 
\end{itemize}

\end{definition}

We define $\Sigma_\emptyset$ as the \emph{empty signature} with $\Sigma_\emptyset \triangleq (\{\},\{\})$, that is, both $F$ and $ar$ are empty sets.

\begin{definition}[Transition System Specifications (TSS)]\label{tss}
A \emph{transition system specification} $\TSS$ is a triple $(\Sigma, L, D)$, where $\Sigma$ is a signature, $\Lab$ is a set of labels, and $D$ is a set of deduction rules. 
We use the symbol $\labelSymbol$ for labels. 
Deduction rules are formed with formulae in the way that we describe below. 
A \emph{positive formula} is of the form $t \labeledStep{\labelSymbol} t'$, and a \emph{negative formula} is of the form $t \noLabeledStep{\labelSymbol}$~~. 
A \emph{formula} $f$ is either a positive formula or a negative formula. 
\emph{Deduction rules} are of the form $(H,f)$, where $H$ is a set of {formulae} called \emph{premises} of the rule, and $f$ is a positive formula called \emph{conclusion} of the rule. 
We write a deduction rule $(H,f)$ as \inference{H}{f}. 
\end{definition}

The notion of derivability of formulae for TSSs with negative premises is from \cite{vanGlabbeek2004}. 
As this definition is standard and we do not use any of its machinery, we do not redefine it, but we write $(\Sigma, L, D) \tssDerive f$ when the formula $f$ is derived from the TSS $(\Sigma, L, D)$ according to the semantics of \cite{vanGlabbeek2004}.

The following definitions from \cite{Groote1992} define the componentwise union of two TSSs. 

\begin{definition}[Union of Signatures]
Given two signatures $\Sigma_1 =  (F_1, \textit{ar}_1)$ and $\Sigma_2 =  (F_2, \textit{ar}_2)$ such that $f\in F_1 \cap F_2 \Rightarrow \textit{ar}_1(f) =\textit{ar}_2(f)$, we have

$\Sigma_1 \oplus \Sigma_2 = (F_1 \cup F_2, \textit{ar}')$, with 
$ \textit{ar}'(f) = \begin{cases} \textit{ar}_1(f), f\in F_1 \\  \textit{ar}_2(f), \textit{otherwise}
 \end{cases}$
\end{definition}

\begin{definition}[Union of TSSs]\label{union}
Given two TSSs $(\Sigma_1, L_1, D_1)$ and $(\Sigma_2, L_2, D_2)$ such that $\Sigma_1 \oplus \Sigma_2$ is defined, we define 
$\unionLanguage {(\Sigma_1, L_1, D_1)}{(\Sigma_2, L_2, D_2)} = (\Sigma_1 \oplus \Sigma_2, L_1 \cup L_2, D_1 \cup D_2)$.
\end{definition}

As an example, we define a TSS for a subset of CCS with inaction \textit{nil}, a unary operator for each action $a$ of a finite set \textit{Act}, and the parallel operator $\parallel$. 
(As usual, \textit{Act} also contains complement actions which can be denoted as $\overline{a}$ for any action $a$.) 
We call this subset \textit{partialCCS}. 
%To avoid confusion with $\lns$ processes of Section \ref{lns}, 
The set of variables $V$ of the TSS of \textit{partialCCS} ranges over $p$, $q$, and so on. 
We define \textit{partialCCS} as follows.  %As usual, transitions are labeled with actions of \textit{Act} or the silent action $\tau$. 

%We define the following set of deduction rules. 

$D\triangleq$
\begin{gather*}
\text{{\large\{}} a.p \labeledStep{a}p, 
\quad 
\inference{p \labeledStep{a}p'}{p \parallel q \labeledStep{a}p' \parallel q}, \quad \inference{q \labeledStep{a}q'}{p \parallel q \labeledStep{a}p \parallel q'},
\\[1ex]
\inference{p \labeledStep{\tau}p'}{p \parallel q \labeledStep{\tau}p' \parallel q},
\quad
\inference{q \labeledStep{\tau}q'}{p \parallel q \labeledStep{\tau}p \parallel q'},
\\[1ex]
\inference{p \labeledStep{a}p' & q \labeledStep{\overline{a}}q'}{p \parallel q \labeledStep{\tau}p' \parallel q'},
\quad
\inference{p \labeledStep{\overline{a}}p' & q \labeledStep{a}q'}{p \parallel q \labeledStep{\tau}p' \parallel q'}\text{{\large\}}}
\end{gather*}

\textit{partialCCS} $\triangleq ((\{\textit{nil},\parallel\} \cup \textit{Act}, ar), \textit{Act} \cup \{\tau\} ,D)$, where $ar$ assigns the arity $0$ to \textit{nil}, the arity $2$ to $\parallel$, and the arity $1$ to every element of \textit{Act}.

\section{Syntax of $\lns$}\label{lns}

The syntax of $\lns$ is defined as follows. % where $l \in \langSet$, $\re\in \regExpVar$, and $w \in \termVarSet$. 
We assume a set of channels $x$, $y$, $z$, and so on. 
We assume that this set and the sets $F$s, $L$s, and $V$s of TSSs (see Definition \ref{signatures} and \ref{tss}) are pairwise disjoint. \\
(Recall that $\TSS$ denotes a TSS, and $t$ is a term. We also use the notation $\many{\cdot}$ for finite sequences.)

\begin{syntax}
   \text{\sf Language Builder} & \lop{\ell} & ::= &  \userLan{\TSS} \mid \lop{\ell \app \unionLop \app \ell} \\
   \text{\sf Regular Expression} & \reBig & ::= &  \constants  \mid \epsilon  \mid  \reBig \concatRE \reBig \mid \reBig ~\mathsmaller{\mid}~ \reBig \mid \kstar{\reBig} \\
   \text{\sf Trace as Reg. Exp.} & \strings & ::= & \constants \mid \epsilon \mid    \strings \concatRE \strings \\
   \text{\sf Transmittable} & e & ::= & x \\
  \textit{(language builders)} 		&&& \mid \userLan{\TSS} \mid e \app \unionLop \app e \\ 
  \textit{(reg. exp.)} 		&&& \mid  \constants  \mid \epsilon  \mid  e \concatRE e \mid e ~\mathsmaller{\mid}~ e \mid \kstar{e} \\ 
  \textit{(terms)} 		&&&  \mid t \\
   \text{\sf Monitor} & \monitor & ::= &  e \Rightarrow P \\
   \text{\sf Process} & P,Q & ::= &\mathbf{0} \mid  x(y).P \mid\overline{x} \langle e\rangle.P  \mid P \parSym Q \mid P + Q  \mid \nu x.P \mid !P \\
  \textit{(online monitoring)}       &&&    \mid  \execLNNMonitor{e}{x}{}{e}{\strings}{}{}{\many{ m}}
		\\
  \textit{(offline monitoring)}       &&&  \mid \isInTrace{e}{e}{P}{Q} \\ % \mid \isInTrace{\strings}{\reBig}{P}{P} \\
  \textit{(checking labels)}		&&&\mid \checkLabels{\many{\constants}}{e}{P}{Q}
\end{syntax}

\emph{Language builder expressions} $\ell$ evaluate to TSSs $\userLan{\TSS}$. 
We can combine two TSSs with \key{union}, which performs the union operation that we have seen in Section \ref{syntaxLanguage}. 

$\lns$ executes programs and keeps track of their execution trace.
We analyze these traces with regular expressions over the set of labels $L$ as the alphabet. 
The grammar of regular expressions $\reBig$ is standard (with $\constants$s as atomic symbols). 
To recall: $\epsilon$ is the empty string, we use an explicit concatenation operator $\concatRE$ (though literature often uses juxtaposition), $\mid$ is the alternation operator, and $\kstar{\reBig}$ is the Kleene closure of $\reBig$. 
As usual, the semantics of a regular expression $\reBig$ is a set of strings. We denote this set with $\langOfRe{\reBig}$. 
The semantics $\langOfRe{\reBig}$ of regular expressions is standard and we omit it here. % \cite{bookAutomata}. 
%The definition of $\langOfRe{\reBig}$ is standard and we omit it here \cite{bookAutomata}. 

Traces are finite strings of labels $\lambda_1\lambda_2 \ldots \lambda_n$. 
We represent traces with regular expressions of the form $\lambda_1\concatRE\lambda_2 \ldots \,\concatRE \lambda_n$, i.e., a concatenation of labels. 
Therefore, we have traces $\strings$ as a special case of regular expressions. The semantics $\langOfRe{\strings}$ of a trace $\strings$ is a singleton set with one string. 

$\lns$ can send and receive \emph{transmittable expressions} $e$ through channels. Transmittable expressions are channels, language builder expressions, regular expressions, and terms. %These syntactic categories can then be bound by names $x$, $y$, and so on. 
Similarly to the $\pi$-calculus, channel names $x$, $y$, and so on, are binding variables for channels. 
Additionally, $\lns$ uses channel names as binding variables for language builder expressions, regular expressions, and terms, as well. 
Transmittable expressions can be language builder expressions that contain variables such as $\TSS \app \unionLop \app x$, where $x$ will be substituted after a communication takes place. 
Similarly, transmittable expressions can be regular expressions that contain variables such as $\reBig \concatRE x$, where $x$ will be substituted later. 
Notice that $\lns$ processes are such that expressions like $\TSS \app \unionLop \app x$ and $\reBig \concatRE x$ will have $x$ already substituted when we reach the moment where these expressions are used. 

$\lns$ contains the processes of the $\pi$-calculus, except that the output prefix sends transmittable expressions. 
Furthermore, $\lns$ contains the following processes. 

$\execLNNMonitor{e_1}{x}{}{e_2}{\strings}{}{}{\many{ m}}$ is a \emph{program execution} with online monitors $\many{ m}$. 
This process evaluates $e_1$ to a TSS $\TSS$. Here, $e_2$ is a term $t$ when this process is activated. 
(We offer some remarks on type errors in Section \ref{discussion}.) 
The term $t$ is the program to be executed. 
This process executes the program $t$ according to the semantics of $\TSS$. 
To do so, we use the derivability of formulae of TSSs to derive a transition from $t$. % one step at a time. 
Program executions evaluate $t$ one step at a time. 
Each of these transitions has a label, and we concatenate these labels in $\strings$. % that is, $\strings$ is an accumulator of the trace of the execution of $t$. 
We assume that every program execution starts with the empty string $\epsilon$.  
Therefore, $\strings$ is the trace of the execution up to a certain point.  
After each transition, we check that the current trace satisfies all the monitors $\many{ m}$. 
Each monitor $m$ contains a regular expression and a process $P$. If the regular expression does not validate the trace then the whole program execution is discarded and $P$ is executed instead. 
If there are multiple monitors that are not satisfied, $\lns$ non-deterministically executes the process of one of them. 
(We purposely under-specify this part. Actual implementations may fix a selection method, for example based on the order in which monitors appear.) 

When the execution of $t$ terminates, the trace $\strings$ is sent over the channel $x$. 

$\lns$ accommodates offline monitors, as well, which analyze the trace of the whole execution after $t$ terminates. 
We do so in the following way. 
As just described, the trace $\strings$ can be received over the channel $x$. 
Afterwards, it can be used with the process $\isInTrace{\strings}{\reBig}{P}{Q}$, which behaves as $P$ if the regular expression $\reBig$ validates the trace $\strings$, 
and behaves as $Q$ otherwise. 
More specifically, our $\isInTraceOp$ operator works in a slightly more general form: $\isInTrace{e_1}{e_2}{P}{Q}$, 
where $e_1$ and $e_2$ are regular expressions $\reBig_1$ and $\reBig_2$ when this process is activated. 
Notice that $\reBig_1$ is not necessarily some trace $\strings$. 
$\isInTraceOp$ checks whether $\reBig_2$ subsumes $\reBig_1$, i.e., $\verification{\reBig_1}{\reBig_2}$
\footnote{The inclusion problem is decidable for regular expressions \cite{inclusionRE}.}.  % \cite{bookAutomata,inclusionRE}
We offer this general form as a convenience to programmers. 
For example, a process may already be planning to run an online monitor with $\reBig_2$, and may receive $\reBig_1$ from another process with instructions to monitor it, as well. 
This process can execute $\isInTrace{\reBig_1}{\reBig_2}{P}{Q}$ and program $P$ to run a monitor with $\reBig_2$ only, as it subsumes $\reBig_1$, as in \\

$
\indent\key{\isInTraceOp}({\reBig_1},{\reBig_2}) ~\texttt{?} ~\execLNNMonitor{\TSS}{x}{}{t}{\epsilon}{}{}{\reBig_2 \Rightarrow \errorExample.\mathbf{0}}\\
\indent\indent\indent\indent\indent\indent\indent\indent \texttt{:}  \\
\indent\indent\indent\indent\indent\indent\indent\indent \execLNNMonitor{\TSS}{x}{}{t}{\epsilon}{}{}{\,\reBig_1 \Rightarrow \errorExample.\mathbf{0}}\\
\indent\indent\indent\indent\indent\indent\indent\indent\indent\indent\indent\indent\indent\indent\indent~~~ {\reBig_2 \Rightarrow \errorExample.\mathbf{0}}
$ \\

Notice that when $\isInTraceOp$ is used with a trace as in $\isInTrace{\strings}{\reBig}{P}{Q}$, then $\langOfRe{\strings}$ is a singleton set with a string and $\verification{\strings}{\reBig}$ holds whenever that string is in $\langOfRe{\reBig}$. 

A process $\checkLabels{\many{\constants}}{e}{P}{Q}$, where $e$ evaluates to a TSS $\TSS$, checks whether the set of labels of $\TSS$ is a subset of the labels $\many{\constants}$. 
In such a case, the process behaves as $P$, otherwise it behaves as $Q$. 
This operation is useful to check, before executing programs, that a TSS works with the expected actions.

\begin{figure}[htbp]
{\small
\textsf{Reduction Semantics}  \hfill \fbox{$P\equiv P$, $P \step P$, \;$\ell \stepL \ell$, \;$P \stepExec P$}
\begin{gather*}
P \parSym \mathbf{0} \equiv P 
\qquad
P \parSym Q \equiv Q \parSym P
\qquad
(P \parSym Q) \parSym R \equiv P \parSym (Q \parSym R)  
\\[1ex]
P + \mathbf{0} \equiv P 
\qquad
P + Q \equiv Q + P
\qquad
(P + Q) + R \equiv P + (Q + R)  
\qquad
!P \equiv P \parSym !P
\\[1ex]
 \nu x.\mathbf{0} \equiv \mathbf{0}
\qquad
 \nu x.\nu y.P \equiv \nu y.\nu x.P
\qquad
 \nu x.(P \parSym Q) \equiv \nu x.P \parSym Q,
~
 \textit{if $x$ is not a free name of $Q$}
\\[1ex]
\inference
	{P_1\step P_1'}
	{P_1+ P_2 \step P_1'}
	\qquad
\inference
	{P_1\step P_1'}
	{P_1\parallel P_2 \step P_1' \parallel P_2}
\qquad
\inference
	{P\step P'}
	{\nu x.P \step  \nu x.P'}
\qquad
\inference
	{P \equiv P' & P' \step Q' & Q' \equiv Q}
	{P \step Q}
\end{gather*}
%\\
{\begin{center} \rule{12cm}{0.5pt} \end{center}}
%\vspace{-3ex}
\begin{gather*}
\ninference{comm}
{}
	{
	x(y).P \parSym \overline{x} \langle e \rangle.Q  \step  P\{e/y\} \parSym Q
}	
\\[2ex]
\ninference{exec}
{
\execLNNMonitor{\TSS}{x}{}{t}{\strings}{}{}{\many{ m}}
\stepExec
P}
{
\execLNNMonitor{\TSS}{x}{}{t}{\strings}{}{}{\many{ m}}
\step
P
}
\quad ~
\ninference{exec-ctx}
{\ell \stepL \ell'}
{
\execLNNMonitor{\userLan{\lop{\ell}}}{x}{\userLan{pn_2}}{\userLan{\LNC{t}}}{\strings}{}{}{\many{ m}} \step \execLNNMonitor{\userLan{\lop{\ell'}}}{x}{\userLan{pn_2}}{\userLan{\LNC{t}}}{\strings}{}{}{\many{ m}}
}
\\[2ex]
\ninference{verify-success}
{\verification{\reBig_1}{\reBig_2}} 
{
\isInTrace{\reBig_1}{\reBig_2}{P}{Q}
\step
P
}
\qquad
\ninference{verify-fail}
{\verificationNO{\reBig_1}{\reBig_2}} 
{
\isInTrace{\reBig_1}{\reBig_2}{P}{Q}
\step
Q
}
\\[2ex]
\ninference{labels-success}
{\TSS= (\Sigma,L,D) \quad \many{\constants} = \constants_1, \cdots, \constants_n \\\\ 
L  \subseteq \{\constants_1, \cdots, \constants_n\}}
{
\checkLabelsShort{\many{\constants}}{\TSS}{P}{Q}
\step
P
}
%\\[2ex]
\quad~~
\ninference{labels-fail}
{\TSS= (\Sigma,L,D) \quad \many{\constants} = \constants_1, \cdots, \constants_n \\\\ 
L  \not\subseteq \{\constants_1, \cdots, \constants_n\}}
{
\checkLabelsShort{\many{\constants}}{\TSS}{P}{Q}
\step
Q
}
\quad~~
\ninference{labels-ctx}
{\ell \stepL \ell'}
{
\checkLabelsShort{\many{\constants}}{\ell}{P}{Q}
\step
\checkLabelsShort{\many{\constants}}{\ell'}{P}{Q}
}
\\[2ex]
\ninference{union}
{}
{\lop{\userLan{\TSS_1}\app \unionLop \app \userLan{\TSS_2}} \stepL \unionLanguage{\userLan{\TSS_1}}{\userLan{\TSS_2}} 
} 
\qquad
\ninference{union-ctx1}
{\ell_1 \stepL \ell_1'}
{{\ell_1}\app \unionLop \app{\ell_2} \stepL {\ell_1'}\app \unionLop \app{\ell_2} 
} 
\qquad
\ninference{union-ctx2}
{\ell_2 \stepL \ell_2'}
{{\TSS}\app \unionLop \app{\ell_2} \stepL {\TSS}\app \unionLop \app{\ell_2'} 
} 
\\[2ex]
\ninference{program-step}
{  {\userLan{\TSS}} \tssDerive  t \labeledStep{\labelSymbol} t'
\\
\many{ m} \equiv  \reBig_1 \Rightarrow P_1 \app \cdots \app  \reBig_n \Rightarrow P_n
\\\\ \langOfRe{\strings\concatRE \labelSymbol} = \{s\}
\\
s\in\langOfRe{\reBig_i} ~~ \textit{for all } 1\leq i\leq n
} 
{
\execLNNMonitor{\TSS}{x}{\userLan{pn_2}}{\userLan{\LNC{t}}}{\strings}{}{}{\many{ m}}
\stepExec 
\execLNNMonitor{\TSS}{x}{}{t'}{\strings\concatRE \labelSymbol}{}{}{\many{ m}}
}
\\[2ex]
\ninference{monitor-fail}
{  {\userLan{\TSS}} \tssDerive  t \labeledStep{\labelSymbol} t'
\\
\many{ m} \equiv  \reBig_1 \Rightarrow P_1 \app \cdots \app  \reBig_n \Rightarrow P_n
\\\\
\langOfRe{\strings\concatRE \labelSymbol} = \{s\}
\\
s\not\in\langOfRe{\reBig_i} ~~\textit{for some } 1\leq i\leq n
} 
{
\execLNNMonitor{\TSS}{x}{\userLan{pn_2}}{\userLan{\LNC{t}}}{\strings}{}{}{\many{ m}}
\stepExec 
P_i
}
%\\[2ex]
\qquad
\ninference{program-end}
{
 {\userLan{\TSS}}  \centernot\tssDerive  t \labeledStep{\labelSymbol} t'
} 
{
\execLNNMonitor{\TSS}{x}{\userLan{pn_2}}{\userLan{\LNC{t}}}{\strings}{}{}{\many{ m}}
\stepExec 
\, !\overline{x} \langle \strings\rangle.\mathbf{0}
}  
\end{gather*}
}
\caption{Reduction semantics of $\lns$. In this figure, \key{lb} is short for the \key{labels} operator.}
\label{fig:dynamicsemantics}
\end{figure}

\section{A Reduction Semantics for $\lns$}\label{operationalSemantics}

Figure \ref{fig:dynamicsemantics} shows the reduction semantics of $\lns$ in two parts. 
The first part of Figure \ref{fig:dynamicsemantics}, that is above the horizontal line, contains the standard definition of the structural congruence $\equiv$ of the $\pi$-calculus, 
and includes the reduction rules of the $\pi$-calculus that are also part of the semantics of $\lns$ \cite{MILNER19921}. %MILNER199241
The second part of Figure 1, that is below the horizontal line, contains the rest of the reduction semantics. 

The main reduction relation is $\step$. 
As in \cite{lns}, this relation makes use of two auxiliary relations: $\stepL$ evaluates language builder expressions $\ell$ into TSSs, 
and $\stepExec$ handles program executions. 

Rule \ruletag{comm} realizes the communication of transmittable expressions.  
Substitution $P\{e/y\}$ substitutes the free occurrences of $y$ in $P$ with $e$. 
This substitution is capture-avoiding, its definition is straightforward, and therefore we do not show it. 
Notice that $\lns$ adopts a call-by-name style for transmitting language fragments. 
In particular, an output prefix $\overline{x} \langle \TSS_1 \app \unionLop \app \TSS_2\rangle.P$ transmits the whole expression $\TSS_1 \app \unionLop \app \TSS_2$ without evaluating it, as it will be evaluated when it is used. 
(The evaluation strategy does not affect the type of scenarios that $\lns$ strives to capture. We chose call-by-name to uniformly use \ruletag{comm} for all transmittable expressions and simplify our calculus.)
%because simplifies our calculus.) 
%that \ruletag{comm} can uniformly be used to pass all %transmittable expressions, which simplifies our calculus.)

Rule \ruletag{exec} handles program executions and simply defers to $\stepExec$-transitions. 
Rule \ruletag{exec-ctx} evaluates $\ell$ when it is not a TSS yet. 

Rule \ruletag{verify-success} checks whether $\reBig_2$ subsumes $\reBig_1$ with $\verification{\reBig_1}{\reBig_2}$. 
In that case, the process takes a transition to $P$. 
Rule \ruletag{verify-fail} fires whenever $\reBig_2$ does not subsumes $\reBig_1$, and executes $Q$. 

Rule \ruletag{labels-success} checks whether the labels of the TSS that is given as second argument are from the set of labels given as first argument.  
%of \key{labels} form 
In that case, the process takes a transition to $P$. 
Rule \ruletag{labels-fail} fires whenever that is not the case, and executes $Q$. 
Rule \ruletag{labels-ctx} evaluates $\ell$ when it is not a TSS yet. 

Rule \ruletag{union} performs the union of two TSSs with the operation $\oplus$ defined in Section \ref{syntaxLanguage}. 
Rules \ruletag{union-ctx1} and \ruletag{union-ctx2} evaluate the first and second argument of \key{union}, respectively. 

Rule \ruletag{program-step} handles program executions $\execLNNMonitor{\TSS}{x}{\userLan{pn_2}}{\userLan{\LNC{t}}}{\strings}{}{}{\many{m}}$. 
We use the derivability relation $\tssDerive$ of TSSs to check that a formula $t \labeledStep{\labelSymbol} t'$ is provable for some $t'$ 
and some label $\labelSymbol$. We then check that all the regular expressions of the monitors $\many{m}$ validate the trace up to that point, 
which is $\strings$ with $\labelSymbol$ appended. 
To do so, we first compute the string $s$ of $\strings\concatRE \labelSymbol$ with $\langOfRe{\strings\concatRE \labelSymbol} = \{s\}$. 
(Recall that the semantics of $\strings\concatRE \labelSymbol$ is a set with one string.) 
Then we check that $s$ belongs to the semantics of each $\reBig_i$ of the monitors (with $s\in\langOfRe{\reBig_i}$). 

Rule \ruletag{monitor-fail} is similar to \ruletag{program-step} except that it fires when there exists a regular expression $\reBig_i$ that does not validate the current trace. 
In this case the transition takes a step to the corresponding process $P_i$ specified by the failing monitor. 
Notice that this transition is non-deterministic when there are multiple regular expressions $\reBig_i$ that fail. 

Rule \ruletag{program-end} detects that a step is not provable for $t$. 
Then, the execution of $t$ is terminated. We spawn a replicated output prefix that sends the trace over the channel $x$. 
The reason for replicating this output is that there may be multiple processes that are interested in analyzing the trace, as we shall see in our second example of Section \ref{examples}. 

\section{Examples}\label{examples}

\begin{figure}
\setlength{\parindent}{0.5cm}

$\textit{timeManagementProvider} \triangleq~ \\
\indent\indent\indent !\channel{whatTask}(y).(\overline{\channel{getTimeManagement}}\langle\textit{parallel}\rangle 
\;+\; 
\overline{\channel{getTimeManagement}}\langle\textit{parallel-max-progess}\rangle)
$\\

\textit{server} $\triangleq$\\
\indent\indent $
\begin{array}{ll} 
&!( \channel{task}_1(x).
				\overline{\channel{whatTask}}\langle\channel{task}_1\rangle. 
					\channel{getTimeManagement}(l).
						\execLNN{\textit{almostTPA\app \unionLop \app l}}{x}{}{\textit{tpa\_program}_1}{\epsilon} \\
\nonumber &~~+\\
\nonumber &~~ \,\channel{task}_2(x).
				\overline{\channel{whatTask}}\langle\channel{task}_2\rangle. 
					\channel{getTimeManagement}(l).
						\execLNN{\textit{almostTPA\app \unionLop \app l}}{x}{}{\textit{tpa\_program}_2}{\epsilon})\\
\end{array}
$\\

\indent$\textit{system} \triangleq (\textit{server}  \parSym \textit{timeManagementProvider} \parSym \textit{client}_1 \parSym \textit{client}_2 \app \ldots \app \parSym \textit{client}_n)$\\
\caption{Server decides idle time vs maximal progress (negative premises).}
\label{fig:example1}
\end{figure}

\paragraph{Example 1 (Negative Premises)}

Our first example makes use of the newly-added feature to use negative premises in the context of processes that communicate languages. 
In this example, we have a server that decides whether parallel processes are allowed to spend idle time or whether they must run with maximal progress. 
We define the TSS of a subset of Hennessy and Regan's Process Algebra for Timed Systems (TPA) \cite{tpa}. 
This subset of TPA contains inaction \textit{nil}, unary operators $a.P$ for each of the actions $a$ of a finite set \textit{Act}, and the parallel operator $\parallel$. 
The transitions of TPA are labeled with actions of \textit{Act}, the silent action $\tau$, and the label $\sigma$ for the passing of idle time. 
The transition $P \labeledStep{\sigma}P$ means that the process $P$ spends idle time. 

We define \textit{almostTPA} to be the subset of TPA just described. However, we omit the rule for the passing of idle time for the parallel operator. 
We first define the set of rules $D\tpaMinusSuffix$. %\\
We then define \textit{almostTPA} as an extension of \textit{partialCSS} of Section \ref{syntaxLanguage}. (Recall that $\Sigma_\emptyset$ is the empty signature defined in Section \ref{syntaxLanguage}.)\\

$D\tpaMinusSuffix = \{a.P \labeledStep{\sigma}a.P, ~ \textit{nil}\labeledStep{\sigma}\textit{nil}\}$. %\\

%We then define \textit{almostTPA} as an extension of \textit{partialCSS} of Section \ref{syntaxLanguage}. (Recall that $\Sigma_\emptyset$ is the empty signature defined in Section \ref{syntaxLanguage}.)\\

\textit{almostTPA} $\triangleq \textit{partialCSS} \oplus (\Sigma_\emptyset, \{\sigma\} ,{D\tpaMinusSuffix})$. \\

We can complete \textit{almostTPA} by including a way for time to pass in the context of the parallel operation. 
For example, we can add either of the following rules. 
\begin{gather*}
\ninference{par-idle}{p \labeledStep{\sigma}p' \\ q \labeledStep{\sigma}q'}{p \parallel q \labeledStep{\sigma}p' \parallel q'}
\qquad
\ninference{par-max}{p \labeledStep{\sigma}p' \\ q \labeledStep{\sigma}q' \\ p \parallel q\noLabeledStep{\tau} }{p \parallel q \labeledStep{\sigma}p' \parallel q'}
\end{gather*}

\ruletag{par-idle} lets the two processes spend idle time, if both processes can. 
Conversely, \ruletag{par-max} implements \emph{maximal progress} and allows idle time to pass only so long that the two processes cannot communicate. 
(TPA uses \ruletag{par-max} in \cite{tpa}.)

We define these two rules in the context of empty TSSs, so that we can conveniently use our union operator to add them to \textit{almostTPA}.  \\

\textit{parallel} $\triangleq  (\Sigma_\emptyset,\{\},\{\ruletag{par-idle}\})$

\textit{parallel-max-progess} $\triangleq  (\Sigma_\emptyset,\{\},\{\ruletag{par-max}\})$\\

Figure \ref{fig:example1} shows our example. 
\textit{server} is a server that offers two services, $\textit{task}_1$ and $\textit{task}_2$. 
Upon a request from a client, \textit{server} executes the program $\textit{tpa\_program}_1$ for $\textit{task}_1$, and $\textit{tpa\_program}_2$ for $\textit{task}_2$. 
These are programs of our subset of TPA. 
\textit{server} has limited computational resources, and executing programs in maximal progress mode is computationally expensive. 
Therefore, \textit{server} communicates with another process called \textit{timeManagementProvider} through the channel \textit{whatTask}, and sends the name of the service that has been requested. 
\textit{timeManagementProvider} non-deterministically decides whether \textit{server} should use maximal progress or not (perhaps based on the urgency of the task, as well as other factors). \textit{timeManagementProvider} sends \textit{parallel} or \textit{parallel-max-progess} through the channel \textit{getTimeManagement}. 
In other words, \textit{timeManagementProvider} decides the semantics of the parallel operator, insofar idle time is concerned, that \textit{server} must use. 
Then, \textit{server} completes \textit{almostTPA} with this fragment of TSS before executing the program.

\paragraph{Example 2 (Offline Monitoring)}

Figure \ref{fig:example2} shows an example with offline monitoring. 
Here, \textit{server} is a server that manages files. 
Clients send programs to \textit{server}. 
Clients also send the TSSs with which \textit{server} must execute these programs. 
\textit{server} is capable of receiving TSSs and executing programs with them. 
However, the only actions that \textit{server} supports are the following actions on files: $\openRE$, $\readRE$, $\writeRE$, and $\closeRE$. 
In other words, clients can define any TSS they wish, and any SOS operator they wish. 
Whichever operators they define, however, must compute transitions to open, read, write, and close files only, as these are the only actions that \textit{server} recognizes. 
When the program terminates, the trace is sent over a channel and is available to both server and clients. % have access to it. 

Our example models the scenario in which both client and server are running an offline monitor to analyze the trace of an execution. 
We describe both sides below. 

\textit{server} receives the language $l$ and the program $w$ through the channel \channel{getProgram}. 
(To shorten our notation, \channel{getProgram} sends and receives multiple arguments in polyadic style, though this is shorthand for a sequence of unary input and output prefixes.)  
\textit{server} also receives the id of the client (as a channel name), and a channel $x$ where to send the trace once the execution of $w$ has finished. 
After receiving these arguments, \textit{server} checks that the set of labels of $l$ is formed with the allowed labels. If this check fails, the server signals an error through the channel $\channel{invalid-language}$. 
Otherwise, the server executes $w$. As there are no online monitors, we simply write $\execLNN{l}{x}{}{w}{\epsilon}$. 
The server is interested in analyzing the trace of this execution, and so it receives the trace at $x$ and runs an offline monitor with $\isInTraceOp$. 
The server checks that $w$ has used files correctly, i.e., it has opened a file before read/write operations, and it has closed the file afterwards. 
The correct use of a file is expressed with the regular expression \textit{fileProtocol}. 
As $w$ may have used files multiple times, the server uses $\isInTraceOp$ to check that the trace is accepted by $\textit{fileProtocol}^{*}$ (with Kleene star). 
If this check succeeds then the server ends. Otherwise, the server flags the client as an unreliable programmer using the channel \channel{flagClient}. 

One of the clients, $\textit{client}_1$, is also interested in analyzing the trace of an execution. 
$\textit{client}_1$ verifies that its program has performed exactly one writing operation. 
This is expressed with the regular expression \textit{onlyOneWrite}. 
$\textit{client}_1$ sends a TSS \textit{tss} and a program \textit{tss\_program} (whose details are irrelevant) to the server. 
It also sends its id and a private channel $x$. 
Then, it receives the trace at $x$, and runs an offline monitor with $\isInTraceOp$ to check that the trace is accepted by \textit{onlyOneWrite}.  
If this check succeeds then $\textit{client}_1$ continues as $P$. 
Otherwise, it terminates. 

\begin{figure}
\setlength{\parindent}{0.5cm}

\textit{allowedLabels} $\triangleq \openRE, \readRE, \writeRE,\closeRE$

\textit{fileProtocol} $\triangleq \openRE \concatRE (\readRE \mid \writeRE)^{*} \concatRE \closeRE$\\

\textit{server} $\triangleq ~!(\channel{getProgram}(l, w,id,x). %~!(\channel{getProgram}(l, w,id).
\\ %\key{language}\textendash
\qquad \indent\indent\indent\indent\indent \key{labels}(\textit{allowedLabels},l) ~ \texttt{?} \\%\checkLabels{l}{\textit{allowedLabels}}{}{}. \\
\indent\indent\indent\indent\indent\indent 
							%\nu x.
							(~ \execLNN{l}{x}{}{w}{\epsilon} \parSym x(\mathit{tr}).\isInTrace{\mathit{tr}}{\textit{fileProtocol}^{*}}{\mathbf{0}}{\overline{\channel{flagClient}}\langle id\rangle}~)\\
\indent\indent\indent\indent\indent\indent \texttt{:}  \\
\indent\indent\indent\indent\indent\indent\overline{\channel{invalid-language}})
$\\
\indent\textit{onlyOneWrite} $\triangleq (\openRE \mid \readRE \mid \closeRE)^{*} \concatRE \writeRE \concatRE (\openRE \mid \readRE \mid \closeRE)^{*}$\\
\indent$\textit{client}_1 \triangleq \nu x.(\overline{\channel{getProgram}}\langle\textit{tss},\textit{tss\_program},id,x\rangle ~\parSym~ x(\mathit{tr}).\isInTrace{\mathit{tr}}{\textit{onlyOneWrite}}{P}{\mathbf{0}})$\\
\indent$\textit{system} \triangleq (\textit{server}  \parSym \textit{client}_1 \parSym \textit{client}_2 \app \ldots \app \parSym \textit{client}_n)$\\
\caption{Server checks for the correct use of files (offline monitoring).}
\label{fig:example2}
\end{figure}

\begin{figure}
\setlength{\parindent}{0.5cm}
\textit{allowedLabels} $\triangleq \openRE, \readRE, \writeRE,\closeRE,\sudoRE, 0,1,2, \ldots,9,\deleteRE$

\textit{ordinary} $\triangleq \openRE \mid \readRE \mid \writeRE \mid \closeRE$ %\mid \sudoRE \mid 0 \mid 1 \mid 2 \mid \ldots \mid 9$

\textit{new} $\triangleq \sudoRE \mid 0 \mid 1 \mid 2 \mid \ldots \mid 9 \mid \deleteRE$\\

\textit{passwordManager} $\triangleq ~!(\overline{\channel{getPasswordOfTheDay}}\langle {3}\concatRE {4}\concatRE{5}\concatRE{6} \rangle)$ \\
\indent\textit{server} $\triangleq ~!(\channel{getProgram}(tss, \textit{prg},id).
\\
\indent\indent\indent\indent\indent \key{labels}(\textit{allowedLabels},tss) ~ \texttt{?}  \\%\checkLabels{l}{\textit{allowedLabels}}{}{}. \\
\indent\indent\indent\indent\indent\indent \channel{getPasswordOfTheDay}(\textit{rexp}).\\
	\indent\indent\indent\indent\indent\indent\indent \nu x. (( \execLNNMonitor{tss}{x}{}{\textit{prg}}{\epsilon}{} \\
\indent\indent\indent\indent\indent\indent\indent\indent\indent (\textit{new}^{*} \concatRE \textit{fileProtocol} \concatRE \textit{new}^{*})^{*} \Rightarrow \overline{\channel{flagClient}}\langle id\rangle)\\
\indent\indent\indent\indent\indent\indent\indent\indent\indent (\textit{ordinary}^{*} \concatRE (\sudoRE \concatRE \textit{rexp} \concatRE \deleteRE) \concatRE \textit{ordinary}^{*})^{*} \Rightarrow \overline{\channel{flagClient}}\langle id\rangle) \\
\indent\indent\indent\indent\indent\indent\indent\indent ~\parSym ~ x(\mathit{tr}).\overline{\channel{end}})\\
\indent\indent\indent\indent\indent\indent \texttt{:}  \\
\indent\indent\indent\indent\indent\indent\overline{\channel{invalid-language}})
$\\

\indent$\textit{system} \triangleq (\textit{server}  \parSym \textit{passwordManager} \parSym \textit{client}_1 \parSym \textit{client}_2 \app \ldots \app \parSym \textit{client}_n)$
\caption{Server receives a regular expression to check valid access to $\deleteRE$ (online monitoring).}
\label{fig:example3}
\end{figure}

\paragraph{Example 3 (Online Monitoring and Sending/Receiving of Regular Expressions)}

Figure \ref{fig:example3} shows an example of online monitoring in $\lns$, and also illustrates the sending/receiving of regular expressions over channels. 
This example refines our previous example. 
Here, \textit{server} additionally admits a privileged action on files, $\deleteRE$, which deletes a file. 
However, programs can perform a $\deleteRE$-transition only if they know the \quoting{password of the day} provided by the process \textit{passwordManager}. 
Passwords are numeric. The password of the example in Figure \ref{fig:example3} is $3456$. 
Programs must first announce their intention to use the privileged action with a $\sudoRE$ action. 
Then, they must perform the actions that correspond to the digits of the password. 
In other words, \textit{server} also admits actions $0$, $1$, $2$, $\ldots$, $9$, where, for example, the action $3$ can be interpreted as \quoting{sent 3} or \quoting{pressed 3}. 
Programs can perform $\deleteRE$ after having performed this sequence of actions. 
In our example, the correct sequence of actions for using $\deleteRE$ is $\sudoRE$, $3$, $4$, $5$, $6$, and $\deleteRE$, in this order. 

\textit{server} receives the language $l$, the program $w$, and the client id. 
(Clients are not interested about the trace in this example, and so they do not send the channel $x$ of the previous example). 
The server checks that the set of labels of $l$ is formed with the allowed labels. 
Then, the server receives a regular expression $e$ through the channel \channel{getPasswordOfTheDay}. 
This represents the fragment of a trace that corresponds to the correct sequence of actions that enables $\deleteRE$. 
The regular expression so received is substituted in lieu of $e$, as we shall discuss shortly. 
At this point, the server creates a private channel $x$ and executes the program $w$ giving $x$ as the channel where to receive the final trace. 
The server also specifies two online monitors for this program execution. 
The first monitor performs the check on file operations that we have seen in the previous example, except that the check is performed at each step of the execution. 
Furthermore, the regular expression of the previous example is slightly modified to take into account the new actions of \textit{server}, which may occur before and after \textit{fileProtocol}. 

The second monitor checks that $\deleteRE$ is used properly. The regular expression of this monitor is 
$(\textit{ordinary}^{*} \concatRE (\sudoRE \concatRE {3}\concatRE {4}\concatRE{5}\concatRE{6} \concatRE \deleteRE) \concatRE \textit{ordinary}^{*})^{*}$ 
after $e$ has been substituted, that is, we check that the correct sequence appears within the other actions. 
If this check fails then the client is flagged for knowing the wrong password, or not using the correct protocol. 

Notice that $\sudoRE \concatRE\sudoRE \concatRE {3}\concatRE {4}\concatRE{5}\concatRE{6} \concatRE \deleteRE$, as well as other acceptable sequences, are invalid. 
We believe that the example sufficiently demonstrates our approach even though our regular expressions could be more refined.  

Finally, \textit{server} can detect that all online monitors succeed throughout the execution of $w$ with the input prefix $x(e).\overline{\channel{end}}$. 
This process signals successful termination through the channel \channel{end}.

\section{Discussion}\label{discussion}

\paragraph{Type Errors and Deadlocks} 

The syntax of $\lns$ does not rule out erroneous uses of its operators. 
For example, (1)  $\execLNNMonitor{\reBig}{x}{}{t}{\strings}{}{}{\many{ m}}$ and (2) $\isInTrace{\strings}{\TSS}{P}{Q}$ are processes of $\lns$. 
The former attempts to execute a program but a regular expression $\reBig$ is given in lieu of a TSS. 
The latter checks the trace $\strings$ against a TSS rather than a regular expression. 
Processes with these and similar type errors can deadlock. 
For example, none of the reduction rules among \textsc{(exec)}, \textsc{(exec-ctx)}, \textsc{(program-run)}, \textsc{(monitor-fail)} and \textsc{(program-end)} of Figure \ref{fig:dynamicsemantics} applies to (1) because $\reBig$ is not a TSS nor a valid language builder expression. 
Similarly, neither \textsc{(verify-success)} nor \textsc{(verify-fail)} applies to (2) because $\TSS$ is not a regular expression. 

As future work, we would like to design a type system that rejects this kind of type errors. Such a type system would rule out the type of deadlocks just described. 
Notice, however, that even if we eliminated deadlocks that are caused by type errors, $\lns$ processes could deadlock anyway in as much the same way that $\pi$-calculus process can deadlock. 
Another source of deadlocks comes from the fact that the union $\oplus$ is sometimes undefined (see Definition \ref{union} of Section \ref{syntaxLanguage}). 

\paragraph{Implementation Aspects} 

Although we have not implemented $\lns$, we discuss some of the aspects that must be taken into account when implementing $\lns$. 

An important aspect is the implementation of program executions. An implementation must be able to take a term and a TSS, and compute the transitions of the term based on the TSS. This may be done by implementing the execution of inference rule systems, including the unification of formulae with the conclusions of inference rules, as well as all the other aspects of inference systems. A good alternative is to outsource this task and make the implementation interact with an external tool. 
For example, the work in \cite{lnp,SLE2018} presents a functional language with language definitions that are based on logic programming, and executes programs using $\lambda$-prolog (see \cite{CiminiFLOPS2020} for implementation details). 
$\lns$ would need to execute TSSs, instead, and tools such as \cite{LTSworkbench} could be used to derive the transitions of terms and provide them to the implementation. 

Similarly, implementors can either implement their own checkers for regular expressions or interact with one of the several tools for regular expressions. % as external tool.

Passing languages through channels is an interesting aspect of the implementation, as well. 
Implementors may use the datatypes that are available in the programming language being used to implement $\lns$, and use them to represent TSSs. 
They then can use the network primitives of the programming language to send/receive values of this datatype, that is, to send/receive fragments of TSSs.

\paragraph{Comparison with HO$\pi$ and Session Types} 
The higher-order $\pi$-calculus (HO$\pi$) can send/receive processes and execute them, which reminds of the core capability of our calculus. 
It is therefore natural to ask whether HO$\pi$ could be used in lieu of $\lns$. 
If all the TSSs that are intended to be used in a $\lns$ process are the HO$\pi$ or subsets of it then HO$\pi$ captures the sending/receiving/executing part of $\lns$. 
Afterwards, HO$\pi$ would need to record the traces of the process being executed, which may not be difficult to do, and would need to represent regular expressions and solve their acceptance problem, which may be harder to do and may take some serious encoding effort. 
If the TSS that we intend to run is, say, an arbitrary language with an operator $f$, then we first need to prove that HO$\pi$ operations faithfully encode $f$. 
We see two issues here: First, HO$\pi$ may not be able to express every operator that can be defined in a TSS with negative premises. 
Second, translating these operators may be possible only when we know their semantics beforehand. However, new operators and new deduction rules are passed around in $\lns$, and via non-deterministic communications. Therefore, we may only know at run-time, right before executing a program, what TSS has been built. 

The monitoring capabilities of $\lns$ remind of session types, as well. 
In this latter approach, 
a process is assigned a type that can express patterns of communications. Session types can express that the trace of the executed process should respect some regular expression. The main difference between monitoring and session types is that the latter are checked statically (compile-time), and reject processes that violate a specified protocol before running them. Monitors, instead, execute the programs to detect violations. It would be interesting to adopt session types rather than monitors in $\lns$. 
However, we see great challenges with such endeavor as we would need to 1) synthesize a session type system for an arbitrary TSS defined by the user, and 2) translate regular expressions as session types of this type system. The task of 1) is also made harder by the fact that, as we pointed out above, TSSs can be non-deterministically built at run-time.

\paragraph{Ideas for Simplifying $\lns$} 
We have designed $\lns$ with linguistic features in mind that one-to-one match the operations that we intended to add. 
There may be room for simplifying $\lns$. 
For example, $\checkLabels{\many{\constants}}{\TSS}{P}{Q}$ can be seen as a special case of $\isInTraceOp$ where the set of labels of $\TSS$ is translated into a concatenation of  labels, like a trace, and where $\many{\constants}$ is translated into a regular expression that accepts all the permutations of the labels of $\many{\constants}$ and their substrings. 
In turn, offline monitoring could be performed with online monitoring, in principle, for example by sending the trace at the end of an execution to an online monitor that executes a program that replays the trace, while checking for a property. 
This suggests that there may be a calculus with online monitoring only, and perhaps auxiliary operations, where \key{labels} and offline monitoring can be encoded as macros. 
We leave working out the details of such a calculus as future work.

\section{Related Work}\label{relatedWork}

\cite{lns} is a direct related work of $\lns$. We have discussed the differences between this paper and \cite{lns} in Section \ref{introduction} (Introduction). 
Also, \cite{lns} transmits language fragments in call-by-value style while $\lns$ does so in call-by-name style. 
\cite{lns} employs a union operation on languages but this operation is not standard, and it has been specifically devised to apply to the syntax for languages of \cite{lns}. 
Instead, we use the standard $\oplus$ operator on TSSs. 
The examples in this paper showcase the added expressiveness of $\lns$ over the prior work done in \cite{lns}. 

There are several works on runtime monitoring (see \cite{monitorSurvey} for a survey). 
Our paper does \emph{not} offer a new monitoring technique. 
On the contrary, we have taken an existing approach, i.e., monitoring with regular expressions \cite{javaMop,eagleMonitor}, %regexpMonitor
and have integrated it into a calculus with processes that communicate programs, traces, and languages. 

Temporal logics such as LTL, and linear fragments of HML, HML with recursion, and the modal $\mu$-calculus (see \cite{AcetoAFIL19}) can be used in lieu of regular expressions to state properties on traces.  %  \cite{Pnueli77}
Our first draft of $\lns$ had LTL formulae in their finite traces interpretation LTL$_f$ \cite{ltlLess1} in lieu of regular expressions. 
However, regular expressions are more expressive than LTL$_f$ \cite{ltlLess1}, and their formalism is more widely known and used, so we simply chose to use that instead. 
We could use an expressive logic but the goal of this paper is not to use the most powerful logic. Rather, we wanted to demonstrate the type of scenarios that $\lns$ enables with a sufficiently expressive formalism that is also easy to read. In this light, we believe that regular expressions may be a suitable choice. 
As future work, we do plan to integrate more expressive logics in $\lns$ and make more sophisticated examples. 

Works such as \cite{aceto2019,FRANCALANZA2021104704} provide general frameworks for monitoring that is based on operational semantics. Some of these works have also been implemented in the \key{detectEr} tool chain\cite{AttardAAFIL21,AttardF16}. 
It would be interesting to integrate these frameworks in $\lns$ in future work. 

We have offered some discussion about the  higher-order $\pi$-calculus and session types in the previous section. 
Generally speaking, the realm of process calculi is tremendously vast and diverse: Process calculi have been augmented with sophisticated operators, and have been applied to a plethora of domains. 
We are not aware, however, of process calculi where processes send fragments of TSSs or regular expressions through channels.

\section{Conclusion}\label{conclusion}

We have presented $\lns$, an extension of $\lnsOLD$ of \cite{lns}. % (\hspace{1sp}\cite{lns}) 
As $\lnsOLD$, our calculus is tailored to model language-oriented scenarios where processes send and receive language fragments. 
$\lns$ also addresses two limitations of \cite{lns}. We use transition system specifications rather than $\lnsOLD$'s specification syntax that is based on higher-order logic programming. 
This allows $\lns$ to define SOS specifications with negative premises. 
Furthermore, we have added monitoring capabilities based on regular expressions. 
Processes of $\lns$ can also send and receive regular expressions. 

We have presented a reduction semantics for $\lns$, and we have provided examples that demonstrate the type of programming scenarios that $\lns$ captures. 
We believe that the three examples in \cite{lns}, together with the examples in this paper, provide a good idea of the potential uses of $\lns$. 

As future work, we would like to design a type system for $\lns$. 
We also would like to extend $\lns$. 
We plan to integrate methods for the automated analysis of language definitions such as that proposed in \cite{lnc}. 
We also plan to add more operations on TSSs, such as removing rules, and renaming operators. 
We observe that the difference between \ruletag{par-idle} and \ruletag{par-max} is the single premise $P \parallel Q\noLabeledStep{\tau}$. 
It would be interesting to make $\lns$ more fine-grained in its capabilities to communicate fragments of TSSs. 
We plan to add the ability of sending/receiving premises which then can be added to rules. 
With such an addition, our first example could simply work with TPA with \ruletag{par-idle}, and add the negative premise above on the fly to make it become \ruletag{par-max}.  
%$P \parallel Q\noLabeledStep{\tau}$

TSSs do not include syntax for binding (and neither does \cite{lns}). 
We plan to integrate the nominal transition systems of Parrow et al. \cite{nominalTSS} in our calculus, which can accommodate binders in SOS specifications. 
With such an addition, we would like to make examples with the $\pi$-calculus and its variants as TSSs that can be sent/received. 

We also would like to investigate a suitable notion of bisimilarity equivalence for $\lns$. 

\paragraph{Acknowledgements} We are thankful to the anonymous reviewers for their suggestions, which helped improve our paper. 

\bibliographystyle{eptcs}
\bibliography{alllns}

\end{document}

%% file: definitions.tex
\newcommand{\lns}{\lnsOLD^{\texttt{+m}}}
\newcommand{\lnsOLD}{\textsc{Lang-n-Send}}
\newcommand{\channel}[1]{\textit{#1}}

\newcommand{\TSS}{\mathcal{T}}
\newcommand{\tssDerive}{\vdash}

\newcommand{\quoting}[1]{``#1''}

\newcommand{\app}{\;}
\newcommand{\key}[1]{\ensuremath{\mathtt{#1}}}
\newcommand{\LNC}[1]{#1}

\newcommand{\many}[1]{\widetilde{#1}}

\newcommand{\verification}[2]{\langOfRe{#1} \subseteq \langOfRe{#2}}
\newcommand{\verificationNO}[2]{\langOfRe{#1}\not\subseteq  \langOfRe{#2}}

\newcommand{\errorExample}{\overline{\channel{monitor-fail}}}

\newcommand{\checkLabels}[4]{\key{labels}(#1,#2) ~\,\texttt{?}\,~ #3 ~\texttt{:}~  #4}
\newcommand{\checkLabelsShort}[4]{\key{lb}(#1,#2) ~\,\texttt{?}\,~ #3 ~\texttt{:}~  #4}

\newcommand{\varSet}{V}

\newcommand{\parSym}{\parallel}
\newcommand{\langOfRe}[1]{\llbracket #1 \rrbracket}

\newcommand{\openRE}{\key{open}}
\newcommand{\closeRE}{\key{close}}
\newcommand{\readRE}{\key{read}}
\newcommand{\writeRE}{\key{write}}
\newcommand{\sudoRE}{\key{sudo}}

\newcommand{\deleteRE}{\key{delete}}

\newcommand{\labeledStep}[1]{\stackrel{#1}{\step}}
\newcommand{\noLabeledStep}[1]{\stackrel{#1}{\centernot\step}}
\newcommand{\tpaMinusSuffix}{_{\textit{tpa}^{-}}}

\newcommand{\Lab}{\textit{L}}

\newcommand{\reBig}{\mathcal{E}}

\newcommand{\labelSymbol}{\lambda}
\newcommand{\constants}{\labelSymbol}
\newcommand{\concatRE}{\cdot}
\newcommand{\kstar}[1]{#1^{*}}

\newcommand{\strings}{\reBig_{\textit{tr}}}
\newcommand{\monitor}{m}

\newcommand{\isInTraceOLD}[4]{\key{isInTrace}(#1,#2) \Rightarrow #3 ~;~  #4}
\newcommand{\isInTraceOpOLD}{\key{isInTrace}}

\newcommand{\isInTrace}[4]{\key{verifyThis}(#1,#2) ~\,\texttt{?}\,~ #3 ~\texttt{:}~  #4}
\newcommand{\isInTraceOp}{\key{verifyThis}}

\newcommand{\stepL}{\longrightarrow_{\textsf{lan}}}
\newcommand{\stepExec}{\longrightarrow_{\textsf{exe}}}

\newcommand{\execLNN}[5]{\texttt{(}#1,#5\texttt{)}\texttt{>}_{#2}\app #4}

\newcommand{\execLNNMonitor}[6]{\texttt{(}#1,#5\texttt{)}\texttt{>}_{#2}\app #4 ~ \key{with} ~ \key{monitors} ~ #6}

\newcommand{\unionLop}{\key{union}}
\newcommand{\ruletag}[1]{\textsc{(#1)}}

\newcommand{\unionLanguage}[2]{#1 \oplus #2}

\definecolor{magenta(dye)}{rgb}{0.79, 0.08, 0.48}
\definecolor{bondiblue}{rgb}{0.0, 0.58, 0.71}
\definecolor{navyblue}{rgb}{0.0, 0.0, 0.5}
\definecolor{lightskyblue}{rgb}{0.53, 0.81, 0.98}
\newcommand{\lop}[1]{#1}
\newcommand{\userLan}[1]{#1}

\newcommand*\colourcheck[1]{%
  \expandafter\newcommand\csname #1check\endcsname{\textcolor{#1}{\text{\ding{52}}}}%
}
\colourcheck{blue}
\colourcheck{green}
\colourcheck{red}

\usepackage{mathtools}
\usepackage{relsize}

\newcommand{\ninference}[3]{\inferrule[(#1)]{#2}{#3}}

\newcommand{\ba}{\begin{array}}
\newcommand{\ea}{\end{array}}

\newenvironment{syntax}{\[\ba{l@{\;\;}lcl}}{\ea\]}

\newcommand{\step}{\longrightarrow}

\definecolor{lightblue}{rgb}{0.25,0.25,1}

\definecolor{lightgray}{gray}{0.9}
\definecolor{darkergrey}{rgb}{0.75, 0.75, 0.75}